# An improved mathematical model of social group competition

I.D. Breslavsky

*A.N. Podgorny Institute for Problems of Mechanical Engineering, Kharkov, Ukraine*

An improved mathematical model of social group competition is proposed. The utility obtained by a member of a certain group from each other member is assumed to be group size-dependent. Obtained results are close to available census data. It is shown that a significant fraction of population can be affiliated in a group with lower maximal specific utility.

Social processes modeling is practically important and indispensable for future prediction task. In present paper the problem of social group competition is considered. Currently, there are a lot of works in this direction [1]. In particular, the specific question – changes in the number of religious adherents – attracts attention of the researchers [1-3].

Present work focuses on study of the social group competition based on the model from paper [2], but the improved law for specific perceived utility is used. The dependence of specific perceived utility (i.e. the utility obtained from each other member of a group) on group size is taken into account.

## MODEL

In [2] the following equation describing the change in the ratio of the number of groups is used:

$$\frac{dx}{dt} = yP_{yx}(x,u_x) - xP_{xy}(x,u_x), \quad (1)$$

where $P_{yx}(x,u_x)$ is the probability, per unit of time, that an individual converts from group Y to X, $x$ is the fraction of the population adhering to X at time t, $0 \leq u_x \leq 1$ is a measure of X's perceived utility; $y = 1-x$, $u_y = 1-u_x$; $P_{xy}(x,u_x) = P_{yx}(1-x, 1-u_x)$. The function $P_{yx}(x,u_x)$ have been taken in the following form:

$$P_{yx}(x,u_x) = cx^\alpha u_x, \quad (2)$$

where $c$ is the time scale coefficient, $\alpha = 1$. In [2] it is demonstrated that the model (1), despite its simplicity, describes the process of competition under various assumptions about the initial properties of the groups X and Y with sufficient accuracy. That's why the model (1) is used in present work.

The dependence (2) expresses the following idea: the attractiveness of the group X increases by the same amount with each 1% of new adherents, irrespective of the X's size. In present paper another assumption is used. As is easily seen, the considered utility is non-material. This follows from the fact that the total utility perceived by all members of society is the non-conserved quantity, it depends on the society members distribution between groups X and Y. Therefore, it is reasonably to assume that the source of this utility is the relationships between members of the group. The values $u_x$ and $u_y$ are different because the rules of interaction within groups have different advantageousness for adherents.

In this paper the assumption that specific perceived utilities $u_x$ and $u_y$ are constants is not used. The following fact is taken into account instead: the members of a small group are ready to do much more for each other than members of large group. When the vast majority of the society belongs to the group X, belonging to this group becomes "normal" and is taken for granted. As a consequence, the members of the dominant group don't want to do something for people around them just because they have common views. This doesn't change the fact that the rules of interaction within certain group may be more advantageous for the individual than in another group. This situation differs from the situation with the language choice, because benefit from the use of language actually increases monotonically with the number of people using it [4].

Thus, the specific perceived utility depends on the size of the group. The following model is used:

$$P_{yx}(x, u_x(x)) = cxu_x(x), \quad (3)$$

$$u_x(x) = \frac{u_x^{max} f_x(x)}{u_x^{max} f_x(x) + u_y^{max} f_y(y)} = \frac{u_x^{max} f_x(x)}{u_x^{max} f_x(x) + (1 - u_x^{max}) f_y(1-x)}, \quad (4)$$

where $u_x^{max}$ is the specific perceived utility of adhering to the group X, when the size of the group is minimal; $u_y^{max} = 1 - u_x^{max}$; $f_x(x)$, $f_y(y)$ are the monotonically nonincreasing functions. In present study two versions of this functions is analyzed:

$$f_x(x) = 1 - (1 - a_x)x, \quad f_y(y) = 1 - (1 - a_y)y; \quad (5)$$

$$f_x(x) = 1 - (1 - a_x)x^2, \quad f_y(y) = 1 - (1 - a_y)y^2, \quad (6)$$

where $0 \leq a_x \leq 1$ is the coefficient of residual utility, i.e. the value that defines the fraction of $u_x^{max}$ which will receive a member of X from each other member in case of all society members belonging to X; $0 \leq a_y \leq 1$ is the similar value for Y. The open question is which function ((5) or (6)) describes real society properties better? The decrease of the specific utility can be the function of the number of group members (5) or the number of contacts between group members (6). Both (5) and (6) functions are used in the present study. Thus, in addition to a single parameter $u_x$ characterizing the relative utility of an individual belonging to one of two groups in the model (2), we get two more parameters $a_x$ and $a_y$ (we have $u_x = u_x^{max}$ for the model (2) where $f_y(y) = f_x(x) \equiv 1$).

## NUMERICAL EXPERIMENT

In paper [2] it is found that model (1, 2) with $u_x = 0.44$ is in a good agreement with the census data of the number of religiously unaffiliated in the Netherlands. The graph obtained with the help of the model (1, 2) for the parameters $u_x = 0.44$, $c = 0.2$, $x_0 = 0.99$ is shown in Figure 1 by dotted line. The graph obtained from the model (1, 3-5) with the parameters $u_x^{max} = 0.4$, $a_x = 0.2$, $a_y = 0.35$, $c = 0.036$, $x_0 = 0.99$ is also shown in Figure 1 (solid line). As is easily seen, for $y < 0.4$, i.e. in the area in which the statistical data are available [2], both curves are quite close. However, further dynamics differ significantly. Using the model (1, 2) one can obtain the prediction of X vanishing, while the model (1, 3-5) gives stable coexistence with the value of $x \approx 0.268$.

Similarly, in Figure 1 the curve obtained from the model (1, 3, 4, 6) with the parameters $u_x^{max} = 0.3$, $a_x = 0.1$, $a_y = 0.2$ $c = 0.031$, $x_0 = 0.99$ is presented by dash-dot line. Here we also see good agreement for small values of $y$ and qualitative difference in the future. There is equilibrium point $x \approx 0.149$ for these parameters.

Another parameter sets for models (1, 3-5) and (1, 3, 4, 6) which give good fit to census data can be found without any difficulty.

It is important to note that even if the maximal specific utility $u_x^{max}$ and the coefficient of residual utility $a_x$ is lower for X it doesn't mean that all people eventually will join to Y.

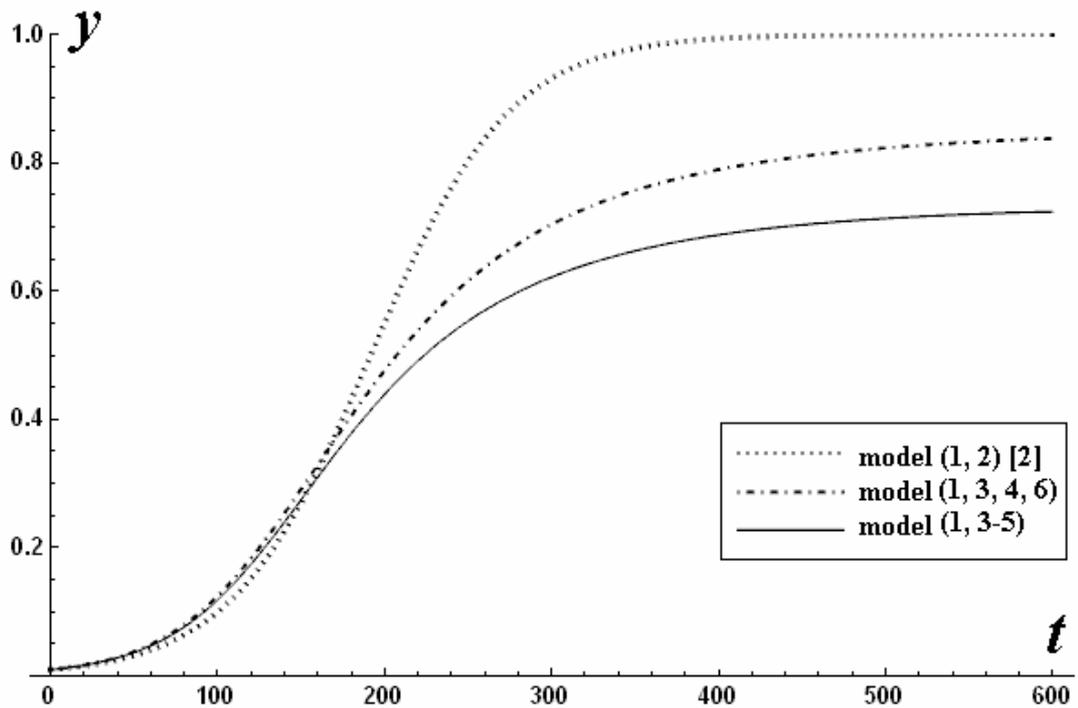

FIG. 1. The growth of percentage religiously unaffliated in the Netherlands versus time. The moment $t=0$ corresponds to 1% of religiously unaffiliated. The census data are available for $y < 0.4$.

## CONCLUSION

The model of competition between social groups is modified in present study. Both the total utility perceived by an individual from group affiliation and the specific utility are assumed to be functions of the size of the group. Results obtained using the proposed model are in good agreement with available statistical data. However, the prediction for future is qualitatively different from that obtained by the model from [2]. To clarify the question which of the models (1, 2), (1, 3-5) or (1, 3, 4, 6) describes better the real competition between social groups, as well as to determine the parameters of these models, large-scale sociological and psychological research are required.

[1] M. Ausloos On religion and language evolutions seen through mathematical and agent based models, arXiv:1103.5382, Proceedings of the First Interdisciplinary CHESS Interactions Conference, C. Rangacharyulu and E. Haven, Eds. (World Scientific, Singapore, 2010) 157.
[2] D.M. Abrams, H.A. Yaple, R.J. Wiener A mathematical model of social group competition with application to the growth of religious non-affliation, arXiv:1012.1375 (2011).
[3] M. Ausloos, F. Petroni Statistical dynamics of religions, Physica A 388 (2009) 4438.
[4] D.M. Abrams, S.H.Strogatz, Modelling the dynamics of language death, Nature 424 (2003) 900.